\documentstyle[preprint,aps,eqsecnum]{revtex}
\tightenlines
\begin{document}
\preprint{SNUTP 97-148}
\draft
\title{
Parametrically amplified radiation 
in a cavity with an oscillating wall
}
\author{Jeong-Young Ji\footnote{Electronic address: jyji@phyb.snu.ac.kr}, 
Kwang-Sup Soh\footnote{Electronic address: kssoh@phya.snu.ac.kr}
}
\address{Department of Physics Education, Seoul National University, 
Seoul 151-742, Korea
}
\maketitle

\begin{abstract}
We introduce a time-dependent perturbation method to calculate the number 
of created particles in a 1D cavity with an oscillating wall of the 
frequency $\Omega . $ This method makes it easy to find the dominant part 
of the solution which results from the parametric resonance. The maximal 
number of particles are created at the mode frequency $\Omega/2 . $ Using 
the Floquet theory, we discuss the long-time behavior of the particle 
creation. 
\end{abstract}

\pacs{03.65.Ca, 42.50.Dv}

\narrowtext
\section{Introduction}

The particle creation by the parametric resonance is a very important 
phenomenon to understand the mechanism of reheating after inflation in the 
early universe~\cite{KofmanLS97}. The photon production by the parametric 
resonance in a cavity with an oscillating wall is another important 
phenomenon to be observed in the experimental situation. Recently, the 
photon creation in an empty cavity with oscillating boundaries has 
attracted much 
attention~\cite{Sassa94,Law94L,MeplanG96,LambrechtJR96}. 
It was also proposed that the high-$Q$ electromagnetic cavities may 
provide a possibility to detect the photons produced in the nonstationary 
Casimir effect~\cite{Dodonov96,DodonovK96}. Therein, they considered the 
vibrating wall with the frequency $\Omega= 2 \omega_{1} $ and found the 
resonance excitation of the electromagnetic modes.

In this paper we consider the quantum (electromagnetic) field in a cavity 
with an oscillating boundary of the frequency $\Omega $. We calculate the 
number of particles (photons) produced by the parametric resonance. For 
the calculation of time-evolution of quantum fields, we introduce a  
time-dependent perturbation method that makes it possible to calculate the 
photon number for any $\Omega $ and to observe clearly the effect of the 
parametric resonance. For a long-time behavior of the quantum field we use 
the Floquet theory.

The organization of this paper is as follows. In Sec.~\ref{sec2} we 
review the scheme of the field quantization in the case of moving 
boundaries. In Sec.~\ref{sec3} we introduce a new perturbation method to 
find the time evolution of the quantum field. Here we write the dominant 
part of the solution of wave equation which results from the parametric 
resonance. We calculate the number of particles created by the vibration 
of the boundary~\cite{J}. In Sec.~\ref{sec4} we use the Floquet theory in 
the perturbation expansion to examine the long-time behavior of the 
solutions. We develope the method to find the characteristic exponent of 
the solution and the periodic part of the corresponding solution that is 
linear combinations of the mode functions. We get the three term 
recurrence relations between the coefficients of the mode functions. The 
last section is devoted to the summary and discussion.

\section{Quantum fields in a 1D cavity with a moving boundary}
\label{sec2}

Let us consider a quantum field obeying the wave equation $(c =1) $
\begin{equation}
\frac{{{\partial^{2}}  A}}{\partial t^{2}} - \frac{{{ \partial^{2}}  
 A}}{\partial x^{2}}=0 
\label{WEQ}
\end{equation}
with time-dependent boundary conditions:
\begin{equation}
A (0, t) = 0 = A ( L(t), t) .
\end{equation}
The field operator in the Heisenberg representation $ A (x, t)  $ can be 
expanded as
\begin{equation}
A (x,t) = \sum_{n} 
[b_{n} \psi_{n} (x,t) + 
b_{n}^{\dagger}\psi_{n}^{*}(x,t) ] ,
\label{Aps}
\end{equation}
where $b_{n}^{\dagger} $ and $b_{n}  $ are the creation and the 
annihilation operators and $\psi_{n} (x,t) $ is the corresponding mode 
function which satisfies the boundary condition 
$\psi_{n} (0, t) = 0 = \psi_{n} ( L(t), t) . $ For an arbitrary moment of 
time, following the approach of Refs.~\cite{Razavy,Calucci,Law94}, we 
expand the mode function as
\begin{equation}
\psi_{n} (x,t) = \sum_{k} Q_{nk} (t) \varphi_{k} (x,t)
\label{psn}
\end{equation}
with the instantaneous basis
\begin{equation}
\varphi_{k} (x, L(t)) = \sqrt{\frac{2}{L(t)}} \sin {\frac{{\pi k x 
 }}{L(t)} } .
\label{phn}
\end{equation}
Here $Q_{nk} (t) $ obeys an infinite set of coupled differential 
equations~\cite{Law95}:
\begin{eqnarray}
{\ddot{Q}}_{nk} + {\omega_{k} }^2 (t) Q_{nk}
&=&  2 \lambda \sum_{j} g_{kj} {\dot{Q}}_{nj}
+ \dot{\lambda} \sum_{j} g_{kj} Q_{nj} 
\nonumber \\
&& + \dot{\lambda}^{2} \sum_{j,l}  g_{jk} g_{jl} Q_{nl}
\label{EOM}
\end{eqnarray}
where $\lambda = \dot{L} / L $ and
\begin{equation}
g_{kj} = { \left\{\begin{array} {cc}(-1)^{k-j} {
\frac{{ 2 k j }}{ j^{2} - k^{2} } }~  & (j \neq k )  
\\ 0  & (j=k) \end{array}\right.} . 
\label{gkj}
\end{equation}
and the time-dependent mode frequency is
\begin{equation}
\omega_{k}  (t) = \frac{{ k \pi }}{L(t)} .
\label{omkt}
\end{equation}

For $L(t \leq 0) = L_{0}  , $ the right hand side of Eq.~(\ref{EOM}) 
vanishes and the solution in this region is chosen to be
\begin{equation}
Q_{nk} (t) = 
\frac{{e^{-i \omega_{k} t}}}{\sqrt{2 \omega_{k}}}
\delta_{nk} .
\label{ICQnk}
\end{equation}
so that the field (\ref{Aps}) can be written as
\begin{equation}
A (x, t \leq 0) = \sum_{n} 
[b_{n} \frac{{e^{-i \omega_{n} t}}}{\sqrt{2 \omega_{n}}} \varphi_{n} (x, 
 L_{0} ) 
+ {\rm H.c. }]
\end{equation}
where $\omega_{n} = \frac{{ \pi n }}{L_{0}} .
 $ For the scalar field or the electromagnetic field, the Hamiltonian 
has the form 
$H = \sum_{n} \omega_{n} ( b_{n}^{\dagger} b_{n} + \frac{1}{2} )  ,
 $ 
and we can interpret $b_{n}^{\dagger} b_{n} $ as the number operator 
associated with the particle with the frequency $\omega_{n} . $

After the change of the boundary, we assume $L (t \geq T) = L_{0} , $ 
then the solution of Eq.~(\ref{EOM}) with the initial condition 
(\ref{ICQnk}), can be written as
\begin{equation}
Q_{nk} (t \geq T) = \alpha_{nk} \frac{{e^{-i \omega_{k} t }}}{\sqrt{2 
 \omega_{k}}} +
\beta_{nk} \frac{{e^{i \omega_{k} t }}}{\sqrt{2 \omega_{k}}} . 
\label{QnkT}
\end{equation}
From (\ref{Aps}) and (\ref{psn}), we have
\begin{equation}
A (x, t \geq T) = \sum_{n} 
[a_{n} \frac{{e^{-i \omega_{n} t}}}{\sqrt{2 \omega_{n}}} \varphi_{n} (x, 
 L_{0} ) 
+ {\rm H.c. }] ,
\end{equation}
where
\begin{eqnarray}
a_{k} &=& \sum_{n} [b_{n} \alpha_{nk} + b_{n}^{\dagger} \beta_{nk}^* ]
\nonumber \\
a_{k}^{\dagger} &=& \sum_{n} [b_{n}^{\dagger} \alpha_{nk}^* 
+b_{n} \beta_{nk} ] . 
\label{a:b}
\end{eqnarray}
Further, it follows from 
$H = \sum_{n} \omega_{n} ( a_{n}^{\dagger} a_{n} + \frac{1}{2} ) 
 $ that $a_{n}^{\dagger} a_{n} $  is the new number operator at 
$t \geq T . $

If we start with a vacuum state $\left| 0_{b} \right> $ such that 
$b \left| 0_{b} \right> = 0 $, 
the expectation value of the new number operator is
\begin{equation}
N_{k} = \left<0_{b} \right| a_{k}^{\dagger} a_{k} \left| 0_{b} \right> 
= \sum_{n=1}^\infty  \left| \beta_{nk} \right|^{2}  ,
\label{NP} 
\end{equation}
which is the number of created particles. 
(Note that the quantum state does not evolve in time in the Heisenberg 
picture.) 

\section{Time evolution of the quantum field
in a cavity with an oscillating boundary}
\label{sec3}

In this section we find the time evolution of quantum field operator 
(\ref{Aps}) by solving Eq.~(\ref{EOM}) with the motion of the wall given 
by
\begin{equation}
L(t) = L_{0} [1 + \epsilon \sin(\Omega t)] .
\label{L}
\end{equation}
Here $\Omega = \gamma \omega_{1} = \gamma \pi / L_{0}  $ and $\epsilon $ 
is a small parameter characterized by the displacement of the wall. For 
$\epsilon \ll  1 , $ having in mind that $\lambda(t) \sim \epsilon $ and 
taking the first order of $\epsilon $ in the mode frequency (\ref{omkt})
\begin{equation}
\omega_{k} (t) = \frac{{ k \pi }}{L_{0}}
[1 + \epsilon \sin(\Omega t)]^{-1} ,
\label{om:e}
\end{equation}
we can replace Eq.~(\ref{EOM}) by a pair of coupled first-order 
differential equations
\begin{eqnarray}
{\dot{Q}}_{nk} &=& P_{nk} 
\nonumber \\
{\dot{P}}_{nk} &=& -{\omega_{k}}^2 ( 1 - 2 \epsilon \sin \Omega t) Q_{nk} 
+ 2 {\dot\frac{{L}}{L} } \sum_{j}  g_{kj} P_{nj} 
\nonumber \\ 
&& + \frac{{\ddot{L} }}{L} \sum_{j}  g_{kj} Q_{nj}  + O (\epsilon^{2} ).
\label{deQP}
\end{eqnarray}

Introducing the new dynamical variables
\begin{equation}
X_{n, k \mp} = \sqrt{ \frac{\omega_{k}}{2}} 
\left( Q_{nk} \pm i \frac{P_{nk}}{\omega_{k}} \right)
\label{X:QP}
\end{equation}
and the vector notation
\begin{equation}
\vec{X}_{n}(t)  = { \left(\begin{array} {c}X_{n,1-} \\ X_{n,1+} \\ 
 X_{n,2-} \\  \vdots \end{array}\right)} ,
\end{equation}
the above equation (\ref{deQP}) can be written as
\begin{equation}
\frac{d}{dt} \vec{X}_{n} (t) = V^{(0)} \vec{X}_{n}(t) 
+ \epsilon V^{(1)} \vec{X}_{n}(t) \label{deXn}
\end{equation}
where $V^{(0)} $ and $V^{(1)} $ are matrices given by
\begin{equation}
V_{k \sigma, j \sigma'}^{(0)} = i \omega_{k} \sigma \delta_{kj} 
 \delta_{\sigma \sigma'}
\end{equation}
and
\begin{equation}
V_{k \sigma, j \sigma'}^{(1)} = \sum_{s = \pm}
\omega_{1} v_{k \sigma, j \sigma'}^{s} e^{s i \gamma \omega_{1} t} ,
\label{Vv}
\end{equation}  
where
\begin{equation}
v_{k \sigma, j \sigma'}^{s} 
= \sigma \gamma g_{kj} \sqrt{\frac{j}{k}} 
\left( \frac{{ \sigma' }}{2} + s \frac{{\gamma}}{4j} \right)
- s \sigma \frac{k}{2} \delta_{kj}
\label{vpm}
\end{equation}
with $s, \sigma,\sigma' = +,-. $ Here we used $\Omega = \gamma \omega_{1} $ 
and $\omega_{k} = k \omega_{1} . $

To find the solution of Eq.~(\ref{deXn}), we introduce a perturbation 
expansion:
\begin{equation}
\vec{X}_{n} =\vec{X}_{n}^{(0)} + \epsilon \vec{X}_{n}^{(1)} 
+ \epsilon^{2} \vec{X}_{n}^{(2)} + \cdots .
\label{series}
\end{equation}
By inserting (\ref{series}) into Eq.~(\ref{deXn}), identifying powers of 
$\epsilon $ yields a series of equations: 
\begin{eqnarray}
\frac{d}{dt} \vec{X}_{n}^{(0)}
&=& V^{(0)} \vec{X}_{n}^{(0)}
\label{0th} \\
\frac{d}{dt} \vec{X}_{n}^{(1)}&=& V^{(1)} \vec{X}_{n}^{(0)} +V^{(0)} 
 \vec{X}_{n}^{(1)} .
\label{1st}
\end{eqnarray}
From the initial condition (\ref{ICQnk}), we have the solution to zeroth 
order equation (\ref{0th})
\begin{equation}
X_{n, k \sigma}^{(0)} = 
\delta_{nk} \delta_{\sigma-} e^{- i \omega_{k} t} ,
\label{0thSol}
\end{equation}
and to the first order equation (\ref{1st})
\begin{equation}
X_{n, k \sigma}^{(1)} (t) 
= \omega_{1} e^{\sigma i k \omega_{1} t} 
\int_{0}^{t} dt' 
v_{k \sigma, n-}^{s} e^{-i(\sigma k - s \gamma +n) \omega_{1} t'} .
\label{solX1}
\end{equation}
When the exponent of exponential function in the integrand of 
(\ref{solX1}) vanishes, we have terms proportional to $\omega_{1} t $ 
which are the effects of parametric resonance. In the usual situation, 
since $\omega_{1} t \gg  1 $, only the resonance terms are dominant and 
the solution becomes by retaining only them:
\begin{eqnarray}
Q_{nk} (t) & \approx &
 \frac{1}{\sqrt{2 \omega_{k}}} e^{-i \omega_{k} t} \delta_{nk} 
\nonumber \\
&+& \frac{{ \epsilon \omega_{1} t}}{\sqrt{2 \omega_{k}}} [
v_{k+,n-}^{+} e^{i \omega_{k} t} \delta_{k, \gamma-n}
\nonumber \\
&+& v_{k-,n-}^{-} e^{-i \omega_{k} t} \delta_{k, n+\gamma}
+ v_{k-,n-}^{+} e^{-i \omega_{k} t} \delta_{k, n-\gamma} ] .
\nonumber \\
&&
\label{QnkReso}
\end{eqnarray}

After some time interval $T$ the wall stops at $x=L_{0} , $ then the 
solution is described by (\ref{QnkT}). By comparing (\ref{QnkT}) with 
(\ref{QnkReso}), the Bogoliubov coefficient $\beta_{nk} $ can be read from 
the solution $Q_{n k} $ to the leading order in $\epsilon $ 
\begin{equation}
\beta_{nk} = \epsilon \omega_{1} T v_{k+,n-}^+ \delta_{k, \gamma-n} ,
\label{bnk}
\end{equation}
which is the coefficient of negative frequency mode function in 
(\ref{QnkReso}).

Using (\ref{gkj}) and (\ref{vpm}), finally we have
\begin{equation}
\left| \beta_{nk} \right|^{2}  =  \frac{1}{4} n k (\epsilon \omega_{1} 
 T)^{2} 
\delta_{k, \gamma-n }  .
\label{be2}         
\end{equation}
Therefore the total number of particles created in the $k $ th mode from 
the empty cavity is
\begin{equation}
N_{k} 
= \sum_{n=1}^\infty \left| \beta_{nk} \right|^{2}  
= { \left\{\begin{array}
{cc}{ \frac{1}{4} } (\gamma-k)k (\epsilon \omega_{1} T)^{2} 
 & k < \gamma \\
0, & {\rm otherwise} .
\end{array}\right.}
\label{Nk} 
\end{equation}
This result is a generalization of Ref.~\cite{DodonovK96} in the short 
time limit $(\epsilon \omega_{1} T \ll  1) $ and it agrees with that 
result for $\gamma =2  $ and $k =1 . $ It should also be noted that the 
maximal number of photons are created at the mode frequency
\begin{equation}
k = \frac{\gamma}{2} ~ {\rm or} ~ \omega_{k} = \frac{\Omega}{2} .
\label{most}
\end{equation}
for $\gamma = {\rm even} $ and at its nearest neighbor frequencies 
$k = (\gamma \pm 1)/2 $ for $\gamma = {\rm odd}. $

\section{Long-time behavior of the solution: 
perturbation approach using Floquet theory}
\label{sec4}
In this section we discuss the long-time behavior of the solution to the 
differential equation (\ref{deXn}). Although the time-dependent 
perturbation method developed in the previous section gives the method to 
calculate the higher order solution, it does not provide the convergency 
of the solution. So it is difficult to examine the long-time behavior of 
the solution. Here we develop another perturbation method using the 
Floquet theory.

Consider a $\tau$-periodic system of differential equations such as 
(\ref{deXn}):
\begin{equation}
\frac{d}{dt} \vec{X} (t) = 
[ V^{(0)} + \epsilon V^{(1)} ] \vec{X} (t) 
\label{deXn1}
\end{equation} 
with the periodic condition
\begin{equation}
[ V^{(0)} + \epsilon V^{(1)} ] (t + \tau)
= [ V^{(0)} + \epsilon V^{(1)} ] (t) ,
\end{equation}
where $\tau = 2 \pi / \Omega . $ The Floquet theory states that the 
solution of (\ref{deXn1}) should be of the form~\cite{Grimshaw}
\begin{equation}
\vec{X} (t) = e^{\mu t} \vec{Z} (t)
\end{equation}
where $\vec{Z} (t) $ is $\tau$-periodic. For simplicity of the indices of 
the matrix, we will slightly change the notation: 
\begin{equation}
\vec{X} (t)  = 
{ \left(\begin{array} {c}\vdots \\ X_{-2}\\ X_{-1}\\ X_{1}\\ X_{2}\\  
 \vdots \end{array}\right)} ,
\end{equation}
\begin{equation}
V_{ kj}^{(0)} = i k \omega_{1} \delta_{kj} ,
\end{equation}
and
\begin{equation}
V_{kj}^{(1)} = \sum_{s = \pm}
\omega_{1} v_{k, j}^{s} e^{s i \gamma \omega_{1} t} ,
\end{equation}  
where
\begin{equation}
v_{k, j}^{s} 
= \gamma g_{kj} \sqrt{\left| \frac{j}{k} \right|} 
\left( \frac{1}{2} + s \frac{{\gamma}}{4j} \right)
- s \frac{k}{2} \delta_{|k|,|j|}
\end{equation}
and $k$ and $j$ are nonzero integers.

To find the solution of Eq.~(\ref{deXn1}), we introduce a perturbation 
expansion:
\begin{eqnarray}
\vec{X} (t) &=& e^{(\epsilon \mu_{1} + \epsilon^{2} \mu_{2} + ...) 
 \omega_{1} t}
\nonumber \\
&& \times
[ \vec{Z}^{(0)} (t) + \epsilon \vec{Z}^{(1)} (t) + \epsilon^{2} 
 \vec{Z}^{(2)} (t) +... ] .
\end{eqnarray}
The zeroth order equation
\begin{equation}
\frac{d}{dt} \vec{Z}^{(0)} = V^{(0)} \vec{Z}^{(0)}
\end{equation}
and the first order equation
\begin{equation}
\frac{d}{dt} \vec{Z}^{(1)} + \mu_{1} \vec{Z}^{(0)} 
= V^{(0)} \vec{Z}^{(1)} + V^{(1)} \vec{Z}^{(0)}
\end{equation}
can be easily solved as:
\begin{equation}
Z_{k}^{(0)} = C_{k} e^{i \omega_{k} t}
\end{equation}
and
\begin{eqnarray}
Z_{k}^{(1)} &=& e^{i k \omega_{1} t}
\int^{t} dt' e^{- i k \omega_{1} t'} 
\nonumber \\
&& \times \sum_{j} 
[ \sum_{s} v_{k, j}^s e^{s i \gamma \omega_{1} t}
 - \mu_{1} \delta_{kj} ]
C_{j} e^{i j \omega_{1} t} .
\label{solZ1}
\end{eqnarray}
When the exponent of the exponential function in the integrand of 
(\ref{solZ1}) vanishes, the integration gives the term proportional to 
$t$. This contradicts the periodicity condition of $\vec{Z} (t) $, 
therefore the coefficient of such term should vanish. Thus, we have the 
following recurrence relation
\begin{equation}
v_{k, k + \gamma}^{-}
C_{k + \gamma} -
\mu_{1} C_{k} + 
v_{k, k - \gamma}^{+}
C_{k - \gamma} 
= 0 .
\label{rr}
\end{equation}
Note that the coefficients are coupled to the $\gamma$th neighbor modes. 
This three term recurrence relation can be written as the following linear 
equation
\begin{equation}
M( \mu_{1} ) \vec{C} = 0
\end{equation}
From the condition for the existence of the nontrivial solution, we can 
find the characteristic exponents by solving
\begin{equation}
{\rm \det} M ( \mu_{1} ) = 0 .
\end{equation}
Then we have the eigenvalues $\mu_{1}^{A} $ and the corresponding 
eigenvectors $\vec{C}^{A} , $ where we introduced the superscript $A$ to 
distinguish the eigenvectors. Then the characteristic solution is 
\begin{equation}
X_{k}^{A} (t \geq 0) = e^{\epsilon \mu_{1}^{A} \omega_{1} t} C_{k}^{A} 
 e^{i k \omega_{1} t} .
\end{equation}
Note that the amplitude (the coefficient of the harmonic function) is 
exponentially increasing when the real part of the characteristic exponent 
$\mu_{1} $ is positive. This is the effect of the parametric resonance. 

By linear combinations of the characteristic solutions, we find the 
solutions 
\begin{equation}
X_{n,k} (t) = \sum_{A} d_{nA} X_{k}^{A} (t) 
\end{equation}
that satisfy the initial conditions
\begin{equation}
X_{n,k} (t \leq 0) 
= e^{i k \omega_{1} t}\delta_{nk} \theta(-n) ,
\end{equation}
where $\theta  $ is the Heaviside unit step function.
From the intial conditions, the coefficients $d_{nA} $ are obtained by 
solving the linear equations
\begin{equation}
\sum_{A} d_{nA} C_{k}^{A} = \delta_{nk} \theta(-n) .
\end{equation}

Since the matrix equation is infinite dimensional, it is difficult to 
find the full solution in a closed form. However, it is helpful to 
consider the truncated matrix in order to understand the long-time 
behavior of the solution. As the simplest model $(\gamma = 2), $ we 
consider only the mode frequency to the next neighboring frequency, e. g. 
to the $\omega_{3} $ for $\omega_{1} $ mode, we have the following 
eigenvalue equation:
\begin{equation}
{ \left(\begin{array}
{cccc}-\mu_{1} & -\sqrt{3}/2 & 0 & 0 \\
\sqrt{3}/2 & -\mu_{1} & -1 & 0 \\
0 & -1 & -\mu_{1} & \sqrt{3}/2 \\
0 & 0 & - \sqrt{3}/2 & -\mu_{1}
\end{array}\right)}
{ \left(\begin{array} {c}C_{-3} \\ C_{-1} \\ C_{1} \\ C_{3} \end{array}\right)}
= 0 ,
\end{equation}
then this equation has the following eigenvalues 
\begin{equation}
\mu_{1} = \pm ( 1 \pm i \sqrt{2} ) /2 
\end{equation}
and the corresponding eigenvectors
\begin{equation}
\vec{C} (\mu_{1} ) = { \left(\begin{array} {c}1 \\ -2 \mu_{1} /\sqrt{3} \\ 
\sqrt{3}/2 + 2 \mu_{1}^{2} / \sqrt{3} \\ \mu_{1} + 4\mu_{1} (\mu_{1}^{2} 
 -1)/3 )\end{array}\right)} .
\end{equation}
In the long-time behavior, the solutions of which characteristic exponent 
is positive will be dominant.

\section{Discussion}

We developed a perturbation method to find the time-evolution of the 
field in a cavity with an oscillating boundary. This method makes it 
possible to calculate the particle number for any oscillation frequency of 
the boundary and to observe clearly the effect of the parametric 
resonance. The results show that the effect of parametric resonance is the 
largest at the half of the frequency of the oscillating 
boundary $( \omega_{k} = \Omega/2 ) . $ This can be understood by 
considering the Mathieu equation
\begin{equation}
\ddot{x} + \mu^{2} ( 1 + \epsilon \cos \Omega t) x = 0 ,
\label{Mathieu}
\end{equation} 
where the parametric resonance takes place most strongly for 
$\Omega = 2 \mu . $
In addition, in the case $\gamma > 2 , $ we see the other resonance 
effects in addition to $\omega_{k} = \Omega/2 , $ which is due to the 
effect of couplings with other mode frequencies in the cavity. 

We used the Floquet theory to examine the long-time behavior of the 
solutions and introduced the general scheme to find the solution that is 
available in the long time. For the simplest case, we found the 
characteristic exponents and the corresponding eigenvectors. In the 
parametric system, the stability-unstability structure is important 
because the resonance condition, $\omega_{n} = \Omega - \omega_{k} $ in 
(\ref{bnk}), is hardly satisfied exactly in the experimental situation. In 
fact the condition of parametric resonance admits some discrepancy as seen 
from the solutions of the Mathieu equation. Therefore it is expected that 
the above simplest case is a good model to study the stability-unstability 
structure in the system of {\it coupled} parametric oscillators. We hope 
to report on this structure in a future paper. Finally we would like to 
mention that it is remained to solve the three term recurrence relation 
(\ref{rr}) in a future study.

\section*{Acknowledgments}
This work was supported by the Center for Theoretical Physics (S.N.U.), 
Korea Research Center for Theoretical Physics and Chemistry, and the Basic 
Science Research Institute Program, Ministry of Education Project No. 
BSRI-96-2418. One of us (JYJ) was supported by Ministry of Education for 
the post-doctorial fellowship.

\end{document}